# Precision frequency comb spectroscopy of the $^{14}N_2O$, $^{14}N^{15}NO$, $^{15}N^{14}NO$, and $^{15}N_2O$ isotopocules in the 3300 – 3550 cm$^{-1}$ range

Adrian Hjältén,[1] Ibrahim Sadiek,[1,2,3] and Aleksandra Foltynowicz[1,*]

[1] Department of Physics, Umeå University, 901 87 Umeå, Sweden
[2] Leibniz Institute for Plasma Science and Technology (INP), 17489 Greifswald, Germany
[3] Experimental Physics V, Faculty of Physics and Astronomy, Ruhr University Bochum, 44780 Bochum, Germany

* aleksandra.foltynowicz@umu.se

**Abstract:**

Nitrous oxide is a long-lived greenhouse gas. Its isotopic composition provides valuable insights into sources and sinks, and about the mechanism of formation. A major challenge in the spectroscopic analysis of the isotopocule compositions is the availability of accurate spectroscopic parameters, particularly for the minor $^{15}N$ isotopocules. In this work, we introduce high-resolution spectroscopic measurements of four isotopocules of nitrous oxide: $^{14}N_2O$, $^{14}N^{15}NO$, $^{15}N^{14}NO$, and $^{15}N_2O$ in the mid-infrared range of 3300 – 3550 cm$^{-1}$ using a frequency comb-based Fourier transform spectrometer. The nitrous oxide samples were obtained from a chemical synthesis involving acid-catalyzed amine-borane reduction of equimolar amounts of $^{15}N$ isotopically enriched sodium nitrite and $^{14}N$ sodium nitrite. The high-resolution spectra, measured in a temperature-controlled single-pass absorption cell, were used to retrieve line center frequencies and relative intensities for a total of 426 rovibrational transitions of the $v_1 + v_3$ band of the four isotopocules, and of the one order of magnitude weaker $2v_2 + v_3$ and $v_1 + v_2 + v_3 - v_2$ bands in the same spectral region. We compare the determined line center frequencies and relative intensities with spectroscopic parameters available in high-resolution molecular databases. For $^{14}N_2O$, $^{14}N^{15}NO$ and $^{15}N^{14}NO$ we find good agreement with the HITRAN database. The $^{15}N_2O$ isotopocule is missing in HITRAN, and we find that its line center frequencies in the GEISA database, the Institute of Atmospheric Optics database, as well as the Ames-1 line list deviate severely from the comb measurements.

## 1. Introduction

Nitrous oxide, $N_2O$, is a potent greenhouse gas, a major precursor for stratospheric ozone depletion, and a tracer of the nitrogen cycle [1, 2]. Natural sources include mainly biological nitrous oxide production in soils and oceans, whereas anthropogenic sources are dominated by agricultural and industrial contributions, with recent studies indicating a 40% increase in the total annual anthropogenic $N_2O$ emissions in the past four decades (1980–2020) [3]. Moreover, with the new directions towards zero-carbon fuels, such as ammonia, increased emissions of reactive nitrogen oxides as well as $N_2O$ is expected to impact human health and climate [4].

The isotopic composition of $N_2O$ provides valuable insights into its sources and formation mechanisms. Isotopic analyses have demonstrated that the isotopomers $^{14}N^{15}NO$ and $^{15}N^{14}NO$ undergo different fractionation processes [5,6]. This makes their ratios, or site preferences highly informative for determining global $N_2O$ fluxes [5], where the site preference relates to $^{15}N$ being in the central (α isotopomer) or the terminal (β isotopomer) position and is calculated as $SP = ([^{14}N^{15}NO]/[^{15}N^{14}NO] - 1) \times 1000$ ‰. Characteristic site preference values have been reported for many biological sources, e.g., for denitrifying bacteria ($SP \approx -5$‰) [6] and nitrifying bacteria ($SP \approx +30$‰) [7]. Furthermore, correlations between total $^{15}N$ content (so-called $^{15}N$ bulk) and agricultural or natural soil sources have been reported along with site preference to elucidate formation mechanisms of $N_2O$ production [8].

There are two major challenges for further progress in $N_2O$ isotope research. One challenge is the lack of standard reference materials, which makes it difficult for different laboratories to ensure their measurements of $N_2O$ isotopocules are consistent and comparable [9, 10]. To address this challenge, research is ongoing to improve the compatibility of laboratory results by providing a reference material as a community standard [11, 12]. Another challenge is related to the measurement techniques used to characterize isotopic signatures of a source sample (from ocean, or soil), atmospheric sample or even a potential reference material. The most widely used technique for $N_2O$ isotopocules measurements is isotope ratio mass spectrometry (IRMS). This technique, however, provides discrete measurements for individual species and its measurement accuracy is limited by the isotope 'scrambling' fragmentation, i.e. the $NO^+$ fragment ions containing the terminal N atom, rather than the central N attached to the O atom as in the original molecule [13]. Site-specific nitrogen isotope ratio measurements based on IRMS need to be corrected for this phenomenon. Furthermore, the IRMS scrambling behavior can vary with time, and it requires regular calibrations. These difficulties are overcome when using optical methods based on Fourier transform infrared (FTIR) spectroscopy and laser spectroscopy that are inherently selective for the different isotopocules. For example, Esler *et al.* [14] utilized conventional low-resolution (1 $cm^{-1}$) FTIR spectroscopy – based on an incoherent light source – to analyze the fundamental asymmetric stretching band (around 2200 $cm^{-1}$) of the α and β isotopomers of $N_2O$. Additionally, Griffith *et al.* [15] used an FTIR spectrometer with an improved resolution of 0.011 $cm^{-1}$ to validate the tropospheric site preference value of Yoshida and Toyoda [1]. However, as high spectral resolution is a key requirement to reveal the isotopomer signatures, instrumental functions of conventional FTIR spectrometers limit the accuracy for concentration determination for isotopomeric species. This limitation is particularly pronounced at low pressures, where narrow linewidths are preferred to enhance the isotopomer selectivity. Higher spectral resolution is provided by spectroscopic techniques using continuous wave (CW) lasers [16 – 19], and instruments suitable for field-deployed measurements have been developed [20 – 24]. However, the narrow bandwidth CW lasers limit their applicability to measurements of one absorption line for each isotopomer, and hence are more prone to changes in analytical conditions (e.g., gas matrix, temperature, pressure), as their large difference in lower state energy will make them respond differently to changes in these conditions.

An additional challenge for optical techniques is the availability of accurate spectroscopic parameters (line center frequencies, line strengths, and pressure broadening coefficients), particularly for the minor $^{15}N$ isotopocules of $N_2O$. The HITRAN2020 database [25] contains line parameters for the first five most abundant isotopocules of nitrous oxide – namely $^{14}N_2^{16}O$, $^{14}N^{15}N^{16}O$, $^{15}N^{14}N^{16}O$, $^{14}N_2^{18}O$, and $^{14}N_2^{17}O$. Most of these line parameters are based on the molecular constants from the work of Toth [26], which contains line lists for eight isotopocules of $N_2O$, in the range of 500 – 7500 $cm^{-1}$. Note that line parameters for spectral regions < 500 $cm^{-1}$ are also included in the HITRAN2020 database for the first four isotopocules, and for spectral region > 7500 $cm^{-1}$ for the fifth isotopocule. However, despite its atmospheric importance, nitrous oxide is one of the least updated molecules in the HITRAN database since the 2004 release with respect to spectral coverage and the number of assigned transitions. A total of 112444 transitions for the $^{14}N_2^{18}O$ isotopocule were included in the 2016 release, and in the last edition, a total of 199 transitions were added for the main isotopocule, $^{14}N_2^{16}O$. This might explain why nitrous oxide was deemed as the forgotten (atmospheric) gas [27]. The GEISA spectroscopic database [28] also contains the $^{15}N_2O$, $^{14}N^{15}N^{18}O$ and $^{15}N^{14}N^{18}O$ isotopocules in addition to those found in HITRAN, and the parameters are again mostly based on the work of Toth [26]. Very recently, empirical line lists have been introduced based on: (i) rovibrational energy levels using the Measured Active Rotational-Vibrational Energy Levels (Marvel) algorithms [29] for the  major $^{14}N_2O$ isotopocule, (ii) the Nitrous Oxide Spectroscopic Line list (NOSL-296) using the effective operator approach [30], also for the  major isotopocule, and (iii) rovibrational energy levels calculated from an accurate potential energy surface and the '*Ames-1*' dipole moment surface at 296 K for the twelve $N_2O$ isotopocules [31]. Additionally, high-temperature line lists valid up to 2000 K for the major $^{14}N_2^{16}O$ and four singly-substituted minor isotopocules ($^{14}N_2^{17}O$, $^{14}N_2^{18}O$, $^{14}N^{15}N^{16}O$, and $^{15}N^{14}N^{16}O$), have recently been included in the ExoMol database [32]. Further experimental investigations are still required to validate these theoretical models and refine the line list parameters. In particular, the $^{15}N_2O$ isotopocule – ranked sixth in the atmospheric abundance – is still missing in the HITRAN database. Except for the data available in the GEISA database [28] and the Ames-1 line list [31], there exists a global modelling study of the measured line lists of $^{15}N_2O$ within the framework of a polyad model of the effective Hamiltonian by Tashkun *et al.* [33]. In that model, the positions of 423 lines and intensities of 392 lines in the 1227 – 3414 cm⁻¹ range were taken from the Toth [26] line list. Note that line positions of Toth [26] are not purely experimental; they were calculated from fitted spectroscopic constants of the measured bands with uncertainties of the calculated line positions ranging between $0.001 - 0.01$ cm⁻¹ (or $30 - 300$ MHz) and of the line intensities ranging from 5% up to 100% (depending on the band). The model results are also available online at the V.E. Zuev Institute of Atmospheric Optics (IAO – https://spectra.iao.ru) database.

In this work, we used mid-infrared frequency comb Fourier transform spectroscopy [34] to measure high-resolution spectra of nitrous oxide obtained from a chemical synthesis involving ¹⁵N isotopically enriched and normal ¹⁴N precursors. The unique combination of large bandwidth, high spectral resolution, and high frequency accuracy provided by frequency combs make them ideal for high-precision state-resolved measurements of the different isotopocules [35 – 38]. Here, we measured spectra of four isotopocules of $N_2O$ simultaneously in the spectral range of $3300 - 3550$ cm⁻¹ with signal-to-noise-ratio (SNR) of up to 970 in a temperature-controlled single-pass absorption cell. The measured spectra were used to retrieve line positions and relative intensities for 426 rovibrational transitions of the $v_1 + v_3$ combination band for all four isotopocules, as well as the one order of magnitude weaker $2v_2 + v_3$ combination band and the $v_1 + v_2 + v_3 - v_2$ hot band in the same spectral region. The determined line positions and relative intensities were compared with spectroscopic parameters available in the HITRAN for the ¹⁴$N_2O$, ¹⁴N¹⁵NO and ¹⁵N¹⁴NO isotopocules, and in the GEISA, IAO and Ames-1 databases for the ¹⁵$N_2O$ isotopocule.

## 2. Equipment and methods

### 2.1 Sample synthesis

For the preparation of isotopically enriched nitrous oxide, we utilized the acid-catalyzed amine-borane reduction of nitrite [39], according to reaction

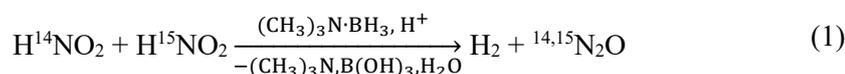

$$H^{14}NO_2 + H^{15}NO_2 \xrightarrow[-(CH_3)_3N,B(OH)_3,H_2O]{(CH_3)_3N\cdot BH_3,\ H^+} H_2 + {}^{14,15}N_2O \qquad (1)$$

Isotopically labeled sodium nitrite salts (Na¹⁵$NO_2$, 98% Sigma Aldrich; Na¹⁴$NO_2$, 98 % Sigma Aldrich) were reduced by stoichiometric amounts of trimethylaminoborane (97%, Sigma Aldrich) in an acidic (pH ~ 0.7) 10% dioxane/water mixture as described by Belly and Kelly [39]. Under such acidic conditions, the reduction of sodium nitrite yields nitroxyl (HNO) or azanone (IUPAC name), which further dimerizes to form $N_2O$. Such a chemical reaction provides a large volume of $N_2O$ in the gas phase suitable for spectroscopic measurements and it is easy to reproduce. We used equal amounts of Na¹⁵$NO_2$ and Na¹⁴$NO_2$ in order to produce four isotopocules of nitrous oxide, namely ¹⁴N¹⁵NO, ¹⁵N¹⁴NO, ¹⁴$N_2O$, and ¹⁵$N_2O$, with similar densities (concentrations), which translates to similar SNR in the spectroscopic measurement.

The setup used for synthesizing and collecting the sample is shown in Figure 1. The reaction was conducted in a spherical glass vial submerged in a water bath of a refrigerated circulator (Thermo Fisher Scientific A10). Nitrogen was bubbled into the solution at a slow rate of $50 - 70$ sccm to strip the formed $N_2O$ in the reaction vessel through an outlet connected to a Teflon tube. The gas flow was then led into a vessel where it was bubbled through a sodium hydroxide (NaOH) solution to capture possible $CO_2$ impurities, before being passed through a dry ice trap to remove water vapor in the gas mixture. Finally, the flow was passed through a U-shaped vial submerged in liquid $N_2$ to condense and trap the isotopically labeled $N_2O$ sample for further spectroscopic analysis.

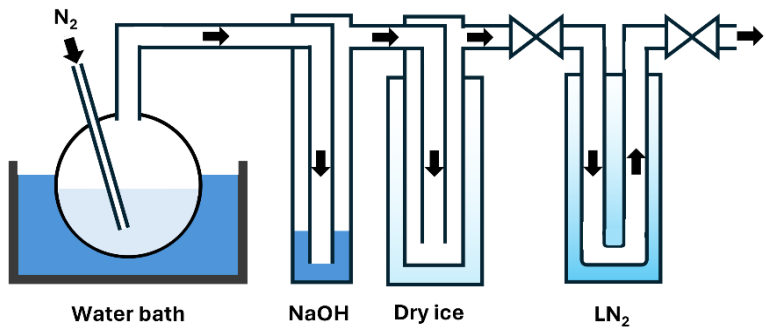

Figure 1. A schematic of the synthesis apparatus. The reaction products are led in a flow of $N_2$ through traps of NaOH to eliminate $CO_2$, dry ice to eliminate $H_2O$, and finally liquid nitrogen ($LN_2$) to trap $N_2O$.

We performed the synthesis at four different temperatures of the water circulator between $278 \pm 1$ K and $336 \pm 1$ K, listed in Table 1, which yields slightly different relative concentrations of the isotopocules. Also listed in Table 1 are the sample pressure, and the number of averaged spectra for each synthesis. Apart from the intended use of the synthesized gas sample for spectroscopic parameter investigations, the synthesis of isotopically enriched $N_2O$ at different temperatures was used to investigate the dimerization of HNO as a potential source of a reference isotopomer ratio standard. Details of the chemical kinetic investigations of the isotopic effect are described in a separate study [40]. For the spectroscopic study in this work, having repeated measurements of the same spectral range on different samples allows reducing the uncertainties on the reported line positions.

Table 1. The absorption measurements performed, indicating the synthesis number that produced the $N_2O$ sample and the synthesis temperature ($T$, in K), the pressure of the $N_2O$ gas used for the spectroscopic measurements ($P$, in mbar), and the number of averaged spectra.

| Synthesis # | $T$ [K] | $P$ [mbar] | Averages |
|---|---|---|---|
| 1 | $296 \pm 1$ | 6.8 | 550 |
| 2 | $316 \pm 1$ | 3.6 | 700 |
| 3 | $336 \pm 1$ | 2.6 | 700 |
| 4 | $278 \pm 1$ | 3.2 | 700 |
| 4 | $278 \pm 1$ | 3.2 | 700 |
| 5 | $316 \pm 1$ | 2.6 | 800 |

## 2.2 Measurement setup

The measurement setup is schematically depicted in Figure 2. It consisted of a mid-IR frequency comb, a temperature-controlled single-pass cell, a fast-scanning Fourier transform spectrometer, and the U-shaped vessel containing the stripped $N_2O$ sample. The frequency comb source has been described previously [34, 41]. Briefly, the offset-frequency-free mid-IR frequency comb was produced by difference frequency generation (DFG) between the high-power 1.030 μm output (pump) of an Yb-doped mode-locked fiber laser (Menlo Systems AB, Orange High Power) with a repetition rate $f_{\text{rep}} = 125$ MHz, and a Raman-shifted soliton (signal) generated from the same source in a highly non-linear fiber, centered at 1.680 μm [42]. The beams were overlapped in a Mg-doped periodically poled lithium niobate crystal (MgO:PPLN) where an idler (the mid-IR comb) was obtained that covered 250 cm$^{-1}$ centered at 3425 cm$^{-1}$. The temporal overlap between the pump and signal pulses in the crystal was maintained by stabilizing the delay line of the pump beam as described in [34]. Two improvements of the comb source were implemented compared to [34]. First, a new low-noise current driver was installed for the pump diode of the Yb-fiber laser oscillator, yielding a factor of 6.5 reduction in intensity noise and thus better SNR in the spectra. Second, the RF-locking of the $f_{rep}$ was improved and simplified by (i) detecting its fourth harmonic at 500 MHz instead of the fundamental frequency and (ii) mixing it with a 500 MHz signal from a single tunable direct digital synthesizer (DDS), instead of two mixing stages as done previously [34]. The DDS was referenced to a GPS-disciplined Rb-clock, and the resulting error signal was fed to a proportional integral controller that acted on an intra-cavity piezoelectric transducer (PZT) in the Yb-fiber laser oscillator. Locking the fourth harmonic instead of the fundamental frequency reduces the comb mode width by a factor of 4, and we measured a width of 2 Hz on the 12[th] harmonic of $f_{rep}$ in closed loop, which corresponds to 140 kHz in the optical domain.

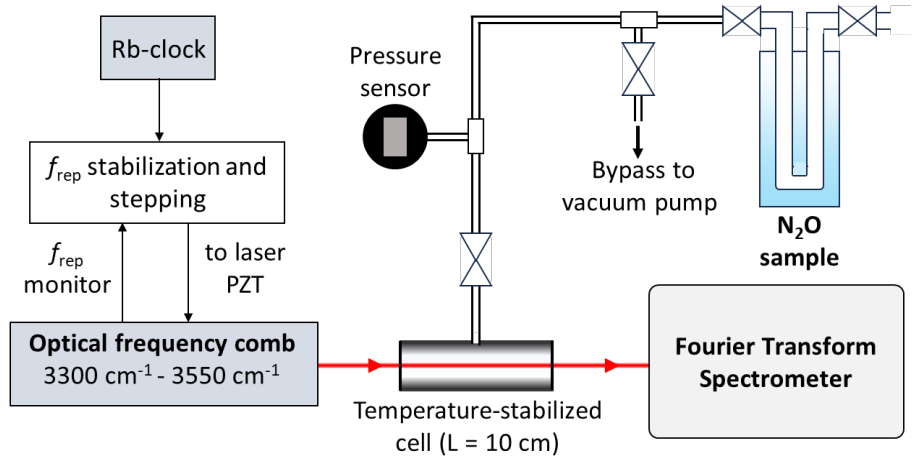

Figure 2. A schematic of the experimental setup for spectroscopy of $N_2O$ isotopocules, including a stabilized mid-IR frequency comb, a temperature-stabilized single-pass cell with a gas supply system for sample delivery and a Fourier transform spectrometer.

The mid-IR beam passed through a 10-cm long Invar single-pass cell connected to a gas supply system. The temperature of the cell was measured using a Pt-100 resistance thermometer and stabilized at 296 K using a Peltier element and a home-written LabVIEW™ program. We estimate the accuracy in temperature to be better than 1 K. The pressure in the gas line leading to the cell was measured using a pressure transducer (CERAVAC CTR 101 N). During the measurements, the cell was sealed off from the rest of the system, and the leak rate was estimated to be below 0.015 mbar/h. The beam transmitted through the cell was directed to the FTS with a nominal resolution matched to $f_{rep}$ of the comb. The optical path difference (OPD) was calibrated using a narrow linewidth 1573 nm CW diode-laser propagating on a path parallel to the comb beam. The FTS was enclosed in a box and purged with dry air to reduce the relative humidity to a few percent during the measurements. This was done to reduce the absorption of the mid-IR light by water in the ambient air and the influence of water dispersion on the calibration of the OPD in the FTS. The comb interferograms were detected in an auto-balanced configuration [43]. Compared to our previous work, we removed the focusing lenses in front of the auto-balanced detector, which reduced the instrumental broadening observed in the spectra (see Section 3.2). The CW reference laser interferogram was acquired with another detector and both signals were recorded by a digitizer (National Instruments, PCI-5922, 5 M sample/s and 20-bit resolution) and collected using a custom LabVIEW™ program.

## 2.3    Spectroscopic measurements

To transfer the produced $N_2O$ sample to the absorption cell, we connected one side of the U-shaped vial to the gas supply system of the spectroscopy setup. To remove the $N_2$ that was still present in the vial at atmospheric pressure, we opened the valve of the vial and slowly reduced the pressure using a vacuum pump to about 3 mbar with the vial still submerged in liquid nitrogen ($LN_2$). We then removed the vial from the $LN_2$ and left it to warm up for a few minutes. Next, we opened the input valve of the absorption cell and filled the cell with the gas while monitoring the absorption of $N_2O$ using the acquisition software of the FTS. We adjusted the pressure in the cell to 2 – 7 mbar to obtain roughly 50% absorption for the strongest absorption lines. Note here that the purity of the sample cannot be known a priori since it stems from a mixture of $N_2O$ and (mainly) $N_2$ in unknown concentrations.

The high-resolution absorption spectra were measured using the sub-nominal resolution sampling-interleaving technique [44, 45]. To this end, we recorded spectra with $f_{rep}$ of the comb locked to four different values separated by 38 Hz, corresponding to shifts of the comb modes of 31 MHz in the optical domain. At each $f_{rep}$ step, we recorded a number of interferograms (550 – 800 depending on the measurement) in series of 50 – 200 interferograms while stepping $f_{rep}$ in alternate directions, and an equal number of background interferograms with the cell evacuated at the first $f_{rep}$ value. Half of these background interferograms were recorded before the absorption measurement and the other half afterwards. Acquiring the background at a single $f_{rep}$ was sufficient since the baseline features were much broader than the comb mode spacing of 125 MHz. The total acquisition time for 800 spectra at each $f_{rep}$ step was 4 h. We Fourier transformed the interferograms and matched the sampling points to the comb mode frequencies as described in [45]. We averaged the sets of spectra at each $f_{rep}$ step and normalized them to the averaged background spectrum, and then calculated the absorption coefficient using the Lambert-Beer law. To remove remaining baseline features arising due to drifts in the comb envelope between the measurement of the absorption and the background spectra, we fitted a sum of a 5th order polynomial and a series of sine terms to account for etalon fringes, and then cancelled the baseline by subtraction. Finally, we interleaved the absorption spectra at the different $f_{rep}$ steps to yield a final spectrum with 31 MHz point spacing. As explained in [45], the sub-nominal resolution method relies on matching the sampling points to the comb mode frequencies when taking the Fourier transform, which requires determining the effective wavelength of the CW reference laser used for OPD calibration. This is done by minimizing instrumental line shape distortions in the interleaved spectra, and here we used the methodology described in [46], which also allows for correcting the nonlinear mapping of the comb frequencies into the FTS spectrum. The uncertainty contributions to line center frequencies stemming from optimizing the reference laser wavelength were estimated as in [46] and were between 1.3 MHz and 2.7 MHz depending on the measurement. These uncertainties are relatively large given the high SNR of the spectra. This is due to the Doppler width of $N_2O$ being comparable to $f_{rep}$, which results in relatively small instrumental line shape distortions, limiting the precision with which the OPD can be calibrated [45].

## 3.    Results

### 3.1    Isotopically-resolved spectra of $N_2O$

Figure 3 shows the interleaved spectrum of nitrous oxide sample synthesized at a temperature of 278(1) K. The spectral range 3300 cm$^{-1}$ – 3550 cm$^{-1}$ covered by the frequency comb is dominated by the $\nu_1$ + $\nu_3$ combination band of all four isotopocules, but it also contains the one order of magnitude weaker $2\nu_2$ + $\nu_3$ combination band and the $\nu_1 + \nu_2 + \nu_3 - \nu_2$ hot band. The inverted spectra show the models of the four isotopocules simulated using the Voigt lines shape with parameters from the HITRAN2020 [25] database for $^{14}N_2O$, $^{15}N^{14}NO$ and $^{14}N^{15}NO$, and from the GEISA database [28] for $^{15}N_2O$. The overall intensity of the model of each isotopocule has been scaled by an arbitrary correction factor to match the measurement, since the absolute concentrations of the four isotopocules are not known. As can be seen in this figure, the model for $^{15}N_2O$ is not complete since GEISA only contains data for the $\nu_1$ + $\nu_3$ band with an intensity cutoff of $1.2 \times 10^{-23}$ cm$^{-1}$/(molecule cm$^{-2}$). Note also the mismatch in intensities between the P- and R-branches (green model) compared to the measurement for this isotopocule. Baseline distortions are present at some positions in the spectra, e.g., the spike around 3510 cm$^{-1}$, and are due to water absorption in the ambient air which was changing during the course of the measurement, and was hence not completely cancelled by background normalization. However, the spikes visible in the middle of the spectrum, e.g., at ~ 3410 cm$^{-1}$, are the result of overlap between lines of different isotopocules. The experimental high-resolution absorption coefficient of $^{14}N_2O$, $^{15}N^{14}NO$, $^{14}N^{15}NO$ and $^{15}N_2NO$ isotopocules is provided as Supplementary Material 1 (SM1).

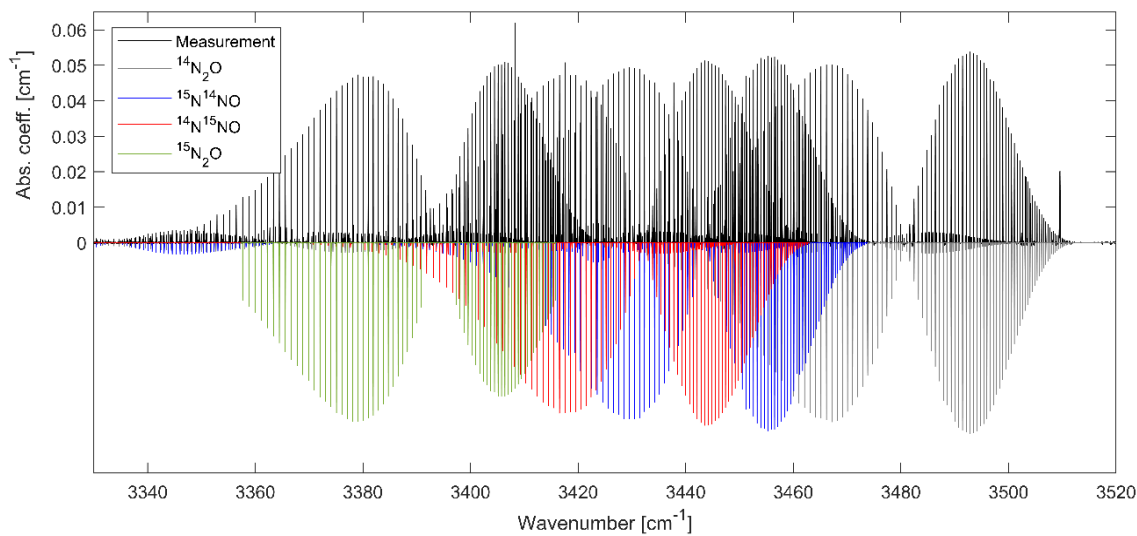

Figure 3. High-resolution absorption coefficient of four isotopocules of $N_2O$ measured at 3.2 mbar (black, pointing up) together with models for $^{14}N_2O$ (grey pointing down, HITRAN), $^{15}N^{14}NO$ (blue, HITRAN), $^{14}N^{15}NO$ (red, HITRAN), and $^{15}N_2O$ (green, GEISA) from a synthesis at a temperature of $278 \pm 1$ K. The model from GEISA contains only a subset of lines of the $\nu_1 + \nu_3$ band.

### 3.2    Line fitting

To retrieve line parameters from the absorption spectra, we fitted Voigt line shapes to lines of all four isotopocules using a home-written MATLAB™ routine based on the Levenberg-Marquardt algorithm. The fitted parameters were the center frequency, the line intensity and the Lorentzian width, where the latter accounted for pressure broadening and possibly remaining minor instrumental broadening. Here, however, the instrumental broadening appeared very small, with the average fitted Lorentzian widths larger by at most 30% (3 MHz) than those predicted by HITRAN for $^{14}N_2O$. This could be within the combined uncertainty of the experimental and predicted widths, especially considering the unknown gas mixture introduced during the synthesis. In any case, the instrumental broadening is significantly reduced compared to that observed in our previous work (28 MHz [34, 41]), which we attribute to better beam quality resulting from the removal of the focusing lenses that were in front of the FTS auto-balanced detector, and the use of a single-pass cell instead of the multi-pass cell as it distorts the beam profile less. The Doppler width was fixed to the theoretical value calculated at room temperature for the listed line centers for each isotopocule. The start values of the parameters were taken from GEISA [28] for $^{15}N_2O$ and from HITRAN2020 [25] for the remaining isotopocules. The sample concentration assumed when fitting was set to the same value (~1.5%) for all isotopocules and chosen to bring the initial model close to the experimental data. Since the GEISA model did not contain all lines, we assigned a number of additional lines of the $\nu_1 + \nu_3$ band using the IAO database and added these to the list of lines to fit. We fitted lines with a predicted line intensity exceeding $1.5 \times 10^{-22}$ cm$^{-1}$/(molecule cm$^{-2}$) (all intensities were rescaled to an abundance of 1) in windows of $\pm770$ MHz corresponding to 4 times the Doppler full-width-at-half-maximum (FWHM) of ~190 MHz. Lines that were separated by 1.5 times the fit window half-width were fitted in a common window, and lines that were closer than 380 MHz were excluded from fitting. Additional lines listed in HITRAN within a fit window that were weaker than the intensity threshold were included in the fit as Voigt line shapes with their parameters fixed to the HITRAN values. Figure 4 shows the absorption profile of the P(26) line of the $\nu_1 + \nu_3$ band of the $^{15}N^{14}NO$ isotopocule (black markers) together with the Voigt fit (red line). The lower panel shows the structureless fit residuals.

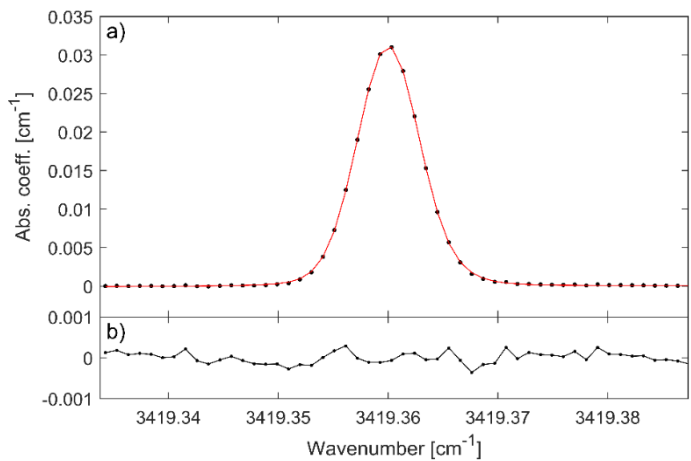

Figure 4. a) The measured P(26) line of $^{15}N^{14}NO$ isotopocule (black markers), and the corresponding Voigt fit (red curve). b) The residuals of the fit.

We applied the fits to transitions in all six spectra, but rejected non-convergent fits and fits with quality factor lower than 30, where the quality factor was defined as the peak absorption of a line divided by the standard deviation of the fit residuals. This resulted in a total of 2474 line fits to 426 distinct transitions of the four isotopocules. Not all these transitions were fitted in all spectra, as some had been rejected according to the criteria described above. Table 2 gives the number of transitions reported here for each isotopocule and vibrational band. For $^{14}N^{15}NO$ and $^{15}N_2O$, the $2\nu_2 + \nu_3$ band is outside of the measured spectral range. For $^{15}N_2O$, 22 of the lines belonging to the $\nu_1+\nu_3$ band reported here are not included in GEISA [28].

Table 2. The number of absorption lines reported in this work for each of the four isotopocules and three vibrational band.

| Isotopocule | $\nu_1+\nu_3$ | $2\nu_2+\nu_3$ | $\nu_1+\nu_2+\nu_3-\nu_2$ |
|---|---|---|---|
| $^{14}N_2O$ | 87 | 11 | 25 |
| $^{15}N^{14}NO$ | 81 | 17 | 24 |
| $^{14}N^{15}NO$ | 77 | - | 24 |
| $^{15}N_2O$ | 80 | - | - |

### 3.2.1 Line positions

We estimated the 1σ uncertainties in line center frequencies as the quadrature sum of the fit uncertainties (570 kHz on average), the uncertainties stemming from the sub-nominal resolution procedure (1.3 to 2.7 MHz, see Section 2.2) and the contribution from the pressure shift. The latter we conservatively estimated to 900 kHz, which is the maximum pressure shift calculated based on the HITRAN and GEISA parameters at the highest measurement pressure of 6.8 mbar. We calculated the mean values of the transition frequencies retrieved from the different spectra, weighted by the inverse square of the uncertainties, and the final uncertainties were obtained by error propagation. These line positions together with their intensities are reported in the Supplementary Material 2 (SM2).

Figure 5a) displays a comparison of experimental line positions for $^{14}N_2O$ (grey), $^{15}N^{14}NO$ (blue) and $^{14}N^{15}NO$ (red) with HITRAN and $^{15}N_2O$ with GEISA (solid green squares) and the IAO line list (open green squares). The inset shows the line center discrepancies for $^{15}N_2O$ with respect to the Ames-1 line list [31], which are larger by one order of magnitude. There is one outlier around 3358 cm$^{-1}$ in the comparison to GEISA (at around +40 MHz) and two outliers at 3358 cm$^{-1}$ and 3417 cm$^{-1}$ in the comparison to IAO (at around +100 MHz), that are out of scale in the plot. The three bands ($\nu_1+\nu_3$, $2\nu_2+\nu_3$ and $\nu_1+\nu_2+\nu_3-\nu_2$) are indicated by the different symbols for each isotopocule, as marked in the legend. The experimental uncertainties vary between 750 kHz and 2.9 MHz and are omitted in the figure for clarity. Some systematic tendencies can be observed in the discrepancies in line center frequencies with respect to HITRAN, though they remain within the upper bound of the HITRAN uncertainty of 30 MHz. The relatively large discrepancies for $^{15}N^{14}NO$ (red, around 3450 cm$^{-1}$) appear to be due to possible perturbation by the $6\nu_2$ state, based on vibrational state energies listed in [26] for the $^{14}N_2O$ isotopocule and assuming similar relative state energies for the $^{15}N^{14}NO$. This perturbation seems not to be properly accounted for in the model of the $^{15}N^{14}NO$ isotopocule. The disagreement for $^{15}N_2O$ is more severe and the models appear to contain systematic errors in the rotational structure as well as an offset in the band centers of about $-17$ MHz for GEISA, 25 MHz for IAO and 740 MHz for Ames-1.

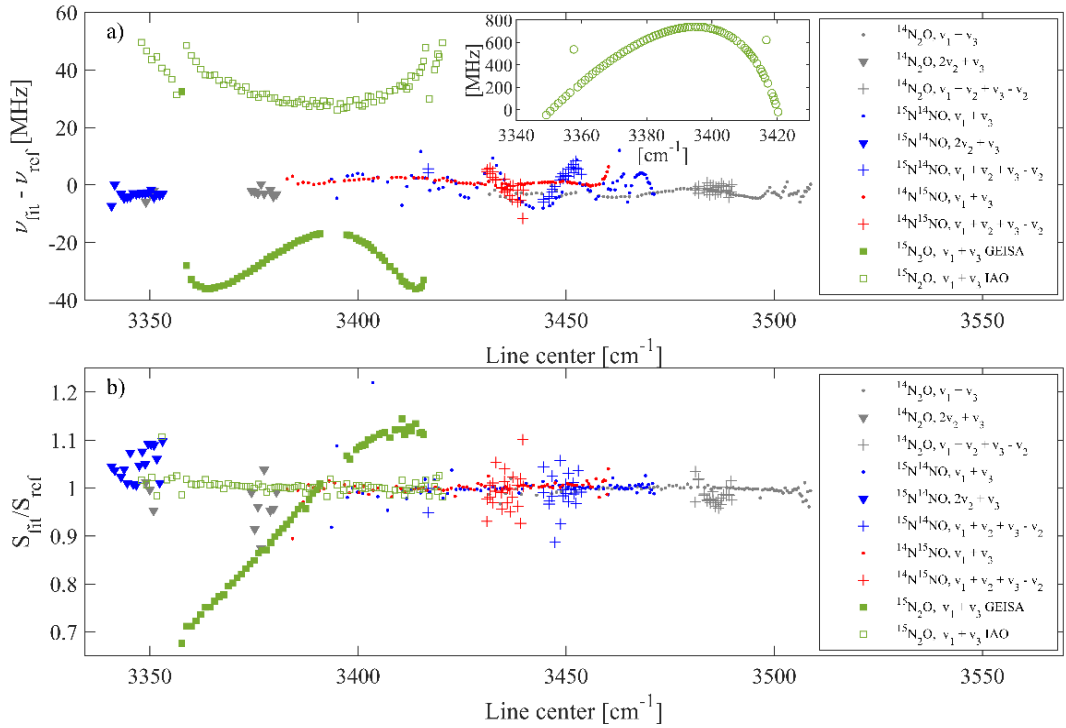

Figure 5. a) Discrepancies between the measured line center positions and the reference data, HITRAN2020 for $^{14}N_2O$ (grey), $^{15}N^{14}NO$ (blue), $^{14}N^{15}NO$ (red) and GEISA (green solid squares) and IAO (green open squares) for $^{15}N_2O$. The inset shows a comparison of $^{15}N_2O$ to the Ames-1 line list. b) The ratios of the experimental intensities to the reference values, where the experimental values were scaled to the listed values using a global constant, are described in text. The $\nu_1+\nu_3$, $2\nu_2 + \nu_3$ and $\nu_1 + \nu_2 + \nu_3 - \nu_2$ bands are denoted by dots, triangles and crosses respectively.

Using the fitted center frequencies of 80 transitions from the P and R branches spanning the entire band, we created a spectral model for the $\nu_1+\nu_3$ band of $^{15}N_2O$ in PGOPHER [47]. Table 3 lists the fit parameters for the upper state, including the band origin, rotational constant $B$, and the centrifugal distortion constants $D$ (quartic) and $H$ (sextic). The lower state parameters were fixed to the experimental values determined from rotational spectroscopy by Bauer *et al*. [48]. We also show the fit parameters obtained from the work of Toth [26] for comparison. By excluding perturbed transitions, namely P(37) and P(38) near 3357.5 cm$^{-1}$, and R(35) and R(36) near 3416.8 cm$^{-1}$, we obtained an average error (observed – calculated) of $6.7 \times 10^{-5}$ cm$^{-1}$ (or 2 MHz), which is within experimental uncertainty of line positions. However, including these perturbed transitions would result in a higher average error of 0.00036 cm$^{-1}$ (or 10.8 MHz). The input PGOPHER file for the $\nu_1+\nu_3$ band of $^{15}N_2O$, using the 80 transitions measured in this work, is provided in the Supplementary Material 3 (SM3).

Table 3. Parameters of the $\nu_1 + \nu_3$ bands of $^{15}N_2O$ obtained from this study together with those obtained by Toth [26]. The lower state constants were fixed to values from rotation spectroscopy [48] in our simulations. Values in parentheses represent the fit parameter uncertainties at 1 $\sigma$ confidence level.

| Constants [cm$^{-1}$] | $10^0 1$ ($\nu_1 + \nu_3$) | | $00^0 0$ ($\nu_0$) |
|---|---|---|---|
| | This work | Toth [26] | Bauer et al. [48] |
| $\tilde{\nu}_0$ | 3394.16240(90) | 3394.16294 | 0.0 |
| $B$ | 0.3998987(50) | 0.39990031 | 0.404860165(28) |
| $D \times 10^7$ | 1.603(7) | 1.618443 | 1.63410(18) |
| $H \times 10^{12}$ | −0.354(200) | -0.2050 | -0.0220(77) |
| Average error | $3.6 \times 10^{-4\,a}$ | | |
| (Obs. − calc.) | $6.7 \times 10^{-5\,b}$ | | |

[a]: including all 80 transitions for that band measured in this work

[b]: excluding perturbed P(37), P(38), R(35) and R(36) transitions

### 3.2.1    Line intensities

Since the purity of the measured samples was unknown and the relative concentrations of the isotopocules varied depending on the synthesis temperature, the retrieved line intensities were not absolute. We found that the synthesis temperature noticeably affected the relative concentrations of the isotopocules, with higher temperature yielding stronger absorption of the heavier isotopocules ($^{15}N_2O$, $^{14}N^{15}NO$, and $^{15}N^{14}NO$) relative to the major $^{14}N_2O$ isotopocule. This higher abundance of the heavier isotopocules compared to the lighter major isotopocule is attributed to the kinetic isotope effect of the HNO dimerization reactions involved in the formation of $N_2O$ [40].

Before averaging the intensities from the six spectra, we scaled the intensities of each isotopocule by a correction factor to match the mean intensities of the respective models. For this, we selected 14 transitions within the P- and R-branches of the $\nu_1 + \nu_3$ band with lower state rotational quantum numbers between 10 and 19. We calculated a correction factor as a ratio of the mean intensities of these lines in the reference model and in the experimental spectra, and we multiplied the latter by this factor. After this rescaling process, we calculated mean intensities for each transition weighted by the relative fit uncertainties. The ratios of the experimental mean intensities of the four isotopocules to those obtained from HITRAN, GEISA and IAO are shown in Figure 5b). Note that a different factor was used to scale the experimental intensities to the IAO predictions than to GEISA, where the intensities in the former were on average larger by 7%.

The relative uncertainties (not shown in the figure) of the reported intensities of all transitions are below 4%. The HITRAN uncertainties of line intensities are <5% and most measured intensities fall within this range. In particular, the residuals of the strong $\nu_1 + \nu_3$ band (grey, blue and red circular markers) relative to HITRAN are flat with no systematic wavenumber dependence, where the standard deviations of the scatter are 1%, 3% and 2% for the $^{14}N_2O$, $^{15}N^{14}NO$ and $^{14}N^{15}NO$ isotopomers, respectively. For a subset of 30 resolved transitions of the $\nu_1 + \nu_3$ band of the α and β isotopomers sharing the same lower state rotational quantum number the relative uncertainties are approximately 0.1%. While these deviations refer to the relative intensities within a given isotopocule, such low relative uncertainties of the intensities of the α and β isotopomers suggest that site-preference measurements for source characterization applications can be performed with a few per mil precision [40]. The spread in relative intensities in the weak vibrational bands: $2\nu_2 + \nu_3$ (black and red downward-pointing triangles for $^{14}N_2O$ and $^{15}N^{14}NO$, respectively) and $\nu_1 + \nu_2 + \nu_3 - \nu_2$ (black, red, and blue plus signs for $^{14}N_2O$, $^{15}N^{14}NO$ and $^{14}N^{15}NO$, respectively) is more pronounced. For the $\nu_1 + \nu_3$ band of the $^{15}N_2O$ isotopocule, the agreement with IAO (green open squares) is comparable to that with HITRAN for other isotopocules, while the discrepancies with GEISA (green solid squares) are much more pronounced, as was noticeable already in Figure 3. The predicted lines of the R-branch are too weak relative to the P-branch by up to 40%.

The relative intensities predicted by Ames-1 line list [31] closely follow those of the IAO line list but are weaker overall by 6%.

## 4.    Conclusions

We used optical frequency comb Fourier transform spectroscopy for high-precision measurements of four isotopocules of nitrous oxide, namely $^{14}N_2O$, $^{14}N^{15}NO$, $^{15}N^{14}NO$, and $^{15}N_2O$ in the 3300 – 3550 $cm^{-1}$ range. We produced nitrous oxide isotopocules from a chemical synthesis involving acid-catalyzed amine-borane reduction of equal amounts of $^{15}N$ and $^{14}N$ isotopically enriched sodium nitrite. Spectra of the gases synthesized at five different temperatures were measured in a temperature-controlled single-pass absorption cell. The final spectra, with signal-to-noise ratio of up to 970, were used to retrieve line positions and relative intensities. Overall, line positions and relative intensities were retrieved for a total of 426 rovibrational transitions of the $v_1 + v_3$ combination band for the four isotopocules, as well as for the one order of magnitude weaker $2v_2 + v_3$ combination band and $v_1 + v_2 + v_3 - v_2$ hot band. We compared the line center positions and relative intensities to those available in the HITRAN2020 and GEISA databases, as well as the line list of Tashkun *et al*. [33], retrieved via the IAO database, and the Ames-1 line list [31]. For the $^{15}N_2O$ isotopologue, the line center frequencies and intensities are missing in the HITRAN database, while they are severely deviating from the comb measurements in the GEISA database. The relative intensities in the IAO and Ames-1 line lists are quite accurate, while the frequencies are again severely deviating from the experiment. The high-precision measurements of the four isotopologues simultaneously using frequency comb Fourier transform spectroscopy help validate theoretical models for line list calculations and fill gaps in current databases. They also serve as a verification of line intensities used for spectroscopic measurements of isotopomer ratios for source characterization applications. The broad bandwidth of frequency combs allows observing state-to-state correlations within the different isotopologues, and hence makes isotopomer ratio measurements less prone to changes in analytical conditions. Moreover, it allows selecting line pairs that are less impaired by potential spectral interfering absorptions.

**Supplementary material**

The supplementary material to this ArXiv submission is available upon request from the corresponding author.

**Acknowledgments**

The authors thank Gernot Friedrichs for support in chemical synthesis, Isak Silander for the design of the single-pass sample cell, Grzegorz Soboń for the loan of the current driver, Iouli E. Gordon for discussions about available spectroscopic data for $^{15}N_2O$, and Kevin K. Lehmann for useful comments on the manuscript.

**Funding**

This project is supported by the Knut and Alice Wallenberg Foundation (KAW 2020.0303) and the Swedish Research Council (2020-00238). I. Sadiek thanks the DFG – Deutsche Forschungsgemeinschaft – for the financial support (project No. SA 4483/1–1).